# A wide-band tunable phase shifter for radio-frequency reflectometry

G. Yin,[1,2] G. A. D. Briggs,[1] and E. A. Laird[1,a]

[1]Department of Materials, University of Oxford, 16 Parks Road, Oxford OX1 3PH, UK
[2]School of Materials Science and Engineering, Tsinghua University, Beijing 100084, China

Radio-frequency reflectometry of nanodevices requires careful separation of signal quadratures to distinguish dissipative and dispersive contributions to the device impedance. A tunable phase shifter for this purpose is described and characterized. The phase shifter, consisting of a varactor-loaded transmission line, has the necessary tuning range combined with acceptable insertion loss across a frequency band 100 MHz – 1 GHz spanning most radio-frequency experiments. Its operation is demonstrated by demodulating separately the signals due to resistance and capacitance changes in a model device.

Radio-frequency (RF) reflectometry is a powerful and versatile experimental tool for measuring nanoscale devices, including single-electron transistors,[1] quantum point contacts,[2,3] spin qubits in quantum dots,[4-6] and nanomechanical resonators.[7] It relies on incorporating the test device into an RF resonant circuit whose reflection coefficient is in general sensitive both to the resistance and capacitance of the device.[8] To identify these contributions, which give different information about the device, it is crucial to separate accurately the two quadratures of the reflected signal.[9,10] This is achieved by applying a tunable phase shift within a homodyne demodulation circuit.

The phase shifter circuit must satisfy several criteria. To isolate either quadrature of the signal, it must have a range of at least 90°. Because the optimum carrier frequency varies between devices and may change for a single device as a function of tuning, magnetic field, or temperature, the ideal phase shifter should work across the range 100 MHz – 1 GHz where most experiments are performed.[2,3,5,6,8,10-12] For good control of the applied power, the insertion loss must not be too large, and should not vary steeply with carrier frequency or control voltage. Finally, internal nonlinearities should be small to prevent spurious signal mixing when frequency multiplexing is used.[11,12]

Here, we describe and characterize a voltage-tunable phase shifter based on a varactor-loaded transmission line and working across the entire desired frequency range. Being made from surface-mount components on a printed circuit board, the design can be readily reproduced and adapted. Previous loaded-line phase shifters have been reported at microwave and radio frequencies using monolithic fabricated circuits.[13,14] Commercially available tunable phase shifters, using varactor or ferrite technology, are usually limited to a frequency range at most one octave.

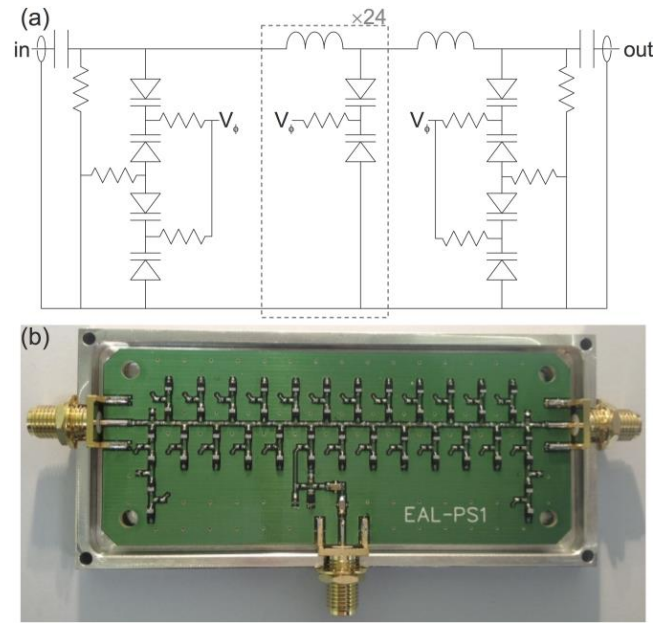

FIG. 1. (Color online). (a) Circuit diagram of the phase shifter. The design consists of a loaded transmission line composed of a series of LC sections, with L = 6.8 nH. The capacitance is provided by voltage-tunable varactors with range ~ 1-9 pF, arranged back-to-back. To improve impedance matching, the end varactors are placed in a doubled configuration. The varactors are tuned using control voltage $V_\phi$, supplied via 20 kΩ resistors. (b) Photograph of the assembled circuit.

The circuit (Fig. 1a) consists of a ladder of inductors of fixed inductance $L = 6.8$ nH and tunable varactors.[15] By applying a control voltage $V_\phi$ between 0 and 30 V, the varactor capacitance $C_V$ can be tuned in the range 1-9 pF. The entire ladder acts as a transmission line of fixed length, whose phase velocity depends on the capacitance of each segment. Varactors are chosen with doping profile such that $C_V \propto V_\phi^{-1/2}$, and arranged back-to-back to suppress nonlinearity.[16] Decoupling capacitors prevent undesired bias through input and output.

Component values are chosen as follows.[13,14] At carrier frequency $\omega/2\pi$ the phase shift per ladder element is

$$\phi_1 \approx \frac{\omega}{2}(C_V Z_0/2 + L/Z_0), \quad (1)$$

and the range of the phase shifter is therefore

$$\Delta\phi = N(\phi_{1,\max} - \phi_{1,\min}) \quad (2)$$
$$\approx N\omega \frac{Z_0}{4}(\sqrt{r} - 1/\sqrt{r})C_{V,0}, \quad (3)$$

where $N$ is the number of elements, $C_{V,0}$ is the geometric center of the varactor range, and $r$ is the ratio of maximum and minimum capacitance. At the lowest frequency of operation, taken as 100 MHz, we require a phase range of at least $\pi/2$, giving

$$\left(\sqrt{r} - \frac{1}{\sqrt{r}}\right)C_{V,0} \gtrsim 8 \text{ pF} \quad (4)$$

with $N = 25$ elements. This is satisfied with $C_{V,0} = 3$ pF and $r = 9$. Matching to the line impedance $Z_0 = 50$ $\Omega$ requires $L \approx C_{V,0} Z_0^2/2 \approx 4$ nH, where $C_{V,0}$ is the geometric center centre of the varactor range. Taking account of simulated parasitic impedances,[17] we choose $L = 6.8$ nH, which gives the necessary range while keeping self-resonances above the maximum operating frequency of 1 GHz.

The phase shifter was characterized using a vector network analyzer[18] (Fig. 2). For different control voltages

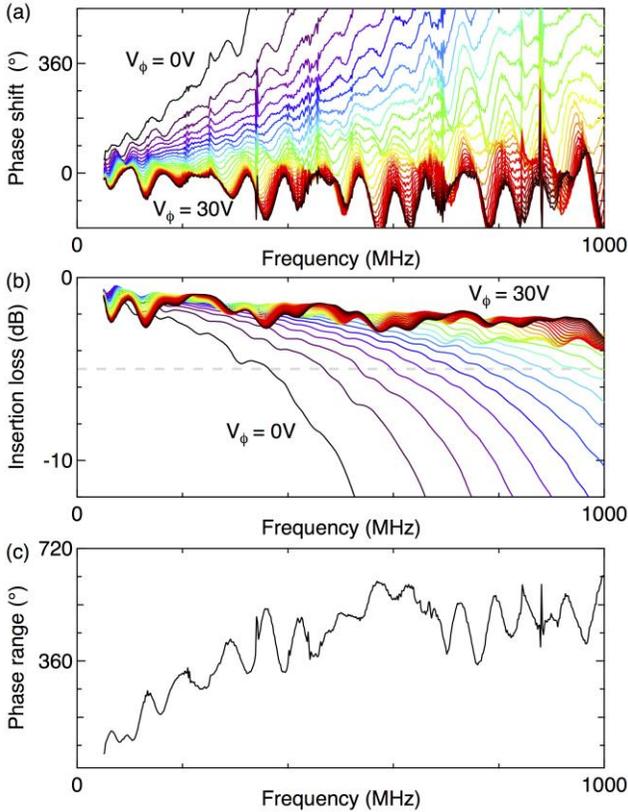

FIG. 2. (Color online). Transmission characteristics of the phase shifter. (a) Phase shift (compared to a fixed delay line) as a function of frequency, plotted for control voltage $V_\phi = 0$ to 30 V in 1 V steps. (b) Insertion loss for the same control voltages. (c) Phase range, defined as the difference of maximum and minimum phase shift at each frequency, for $V_\phi$ values giving insertion loss better than -5 dB (level marked with dashed line in (b)).

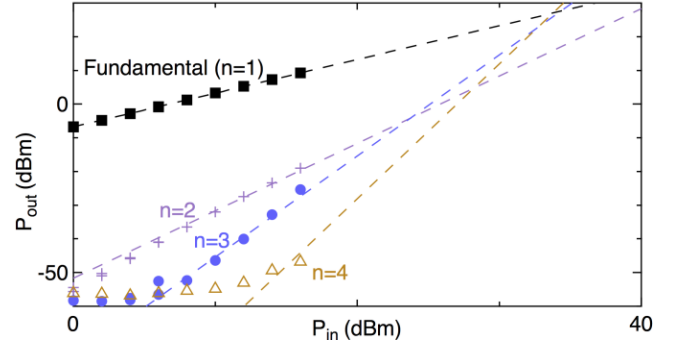

FIG. 3. (Color online). Nonlinear behavior of the phase shifter. Markers: Transmitted power $P_{\text{out}}$ in each harmonic, plotted as a function of input power $P_{\text{in}}$ at 279.33 MHz with $V_\phi = 10$ V. Lines: fits of the form $P_{\text{out}} = AP_{\text{in}}^n$. Extrapolation of the fits gives intercepts IIP2 = 45 dBm, IIP3 ≈ IIP4 = 34 dBm, indicating that harmonic generation is insignificant for typical application power $P_{\text{in}} \lesssim 13$ dBm.

$V_\phi$, the phase and amplitude of the circuit are shown as a function of frequency. As expected, increasing $C_V$ by decreasing $V_\phi$ leads to a larger phase shift (Fig. 2a). At higher frequencies, this becomes more pronounced because the same delay corresponds to a larger phase. The insertion loss also depends on $V_\phi$ (Fig. 2b); at high frequency, the circuit approaches its self-resonance when $C_V$ is tuned near its maximum. Tuning the circuit therefore involves a trade-off, because at high frequency the range of $V_\phi$ must be reduced to obtain acceptable insertion loss. This is quantified in Fig. 2c, which plots the range over which the phase can be tuned while maintaining acceptable insertion loss, defined as -5 dB. The desired tuning range of at least 90° was obtained over the range 60 MHz – 1 GHz.

Nonlinear behavior of the phase shifter is shown in Fig. 3. A carrier signal at 279.33 MHz was applied at the input, and the output was measured with a signal analyzer.[19] Internal mixing due to diode nonlinearity leads to output harmonics, plotted as a function of power. In a reflectometry experiment, these could lead to spurious signals. As shown, these harmonics are negligible at typical powers used in experiments; the power seen in a reflectometry experiment is usually below 13 dBm, set by the mixer level in a demodulation circuit. Extrapolating measured harmonics to high power leads to intercept (IIP$n$) values above 30 dBm, characterizing the power at which nonlinearity would become significant.[20]

The phase shifter is applied to a reflectometry circuit in Fig. 4. In a real nanoscale device, changes in resistance might be detected from a charge sensing single-electron transistor,[1] while changes in capacitance might indicate the spin parity of a two-electron quantum dot.[5] These effects are simulated by a MOSFET in parallel with a 10 pF voltage-controlled varactor,[21] incorporated into a tank circuit resonating at 279.33 MHz. In the measurement configuration of Fig. 4a, a carrier signal at the resonance frequency is sent via a directional coupler to the tank circuit. The reflected signal is passed through the phase shifter and mixed with the carrier to generate an intermediate-frequency signal proportional to the

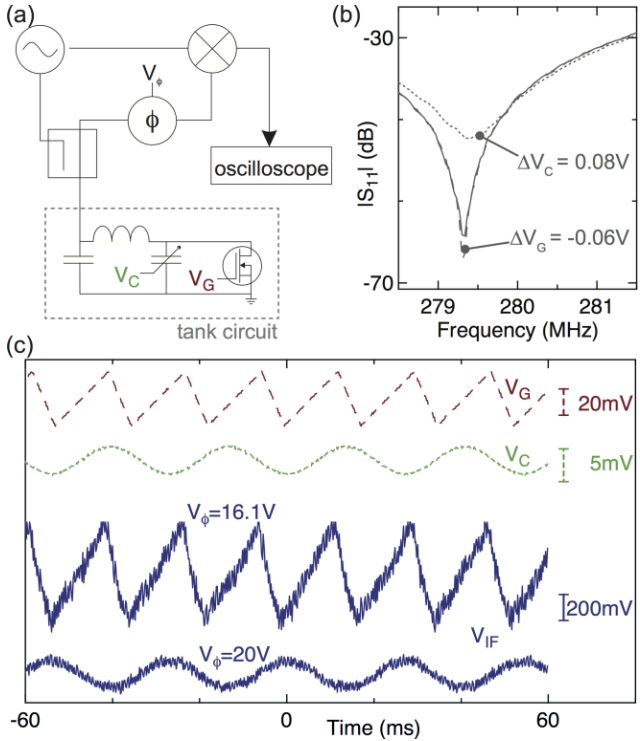

FIG. 4. (Color online). Demonstration of the phase shifter to separate real and complex impedance signals. (a) Simplified measurement setup. A carrier signal is split, with part of the signal sent via a directional coupler to an LCR tank circuit. The reflected signal is passed through the phase shifter and mixed with the original carrier to generate the detected signal $V_{IF}$. The capacitance and resistance of the tank circuit are tuned via $V_C$ and $V_G$ respectively. (b) Reflectance from tank circuit as a function of frequency with $V_G$ = -2.2 V, $V_C$ = 7.9 V (solid line), and with small voltage changes applied to $V_G$ and $V_C$ (dashed/dotted lines). The tank resonance is evident as a dip in $|S_{11}|$. Whereas adjusting $V_G$ predominantly changes the damping and therefore the depth of the resonance, adjusting $V_C$ also shifts its frequency. (c) Top traces: time-dependent voltages applied simultaneously to $V_G$ (dashed trace) and $V_C$ (dotted trace). The corresponding demodulated signal $V_{IF}$ (solid traces) can be tuned using the phase shifter to be predominantly sensitive to the amplitude ($V_\phi$ = 16.1 V) or to the phase ($V_\phi$ = 20 V) of the reflected signal, thereby distinguishing the separate effects of $V_G$ and $V_C$.

reflectance of the tank circuit, and therefore dependent on the impedance of the device.[2]

A resonance can be seen as a dip in the reflection coefficient $|S_{11}|$ of the tank circuit as a function of frequency (Fig. 4b). The separate effects of tuning resistance and capacitance can be seen by adjusting the gate voltage $V_G$ and varactor voltage $V_C$. Whereas a change in resistance leads to a change predominantly in the quality factor of the resonance, a change in capacitance also affects the center frequency.

To demonstrate that the phase shifter allows the demodulation circuit to separate these effects, we apply separate excitations to $V_G$ and $V_C$ (Fig. 4c). In a simple LCR resonator, changes in resistance affect the amplitude of the reflected signal, while changes in capacitance affect its phase. These are evident in the in-phase and quadrature components of the demodulated signal $V_{IF}$. (In this circuit, the two components are not precisely orthogonal because of the fixed impedance-matching capacitor shown in Fig. 4a, and because of the changing capacitance of the MOSFET.) By tuning the phase control voltage, we can choose to detect either component of the detected signal, showing that we can distinguish separate changes in the complex impedance of the device. By splitting the reflected signal between two phase shifters tuned 90° apart, it would be possible to monitor the complex impedance of the device in real time. The frequency range of the phase shifter could be changed by incorporating different inductors or a different number of segments.

We acknowledge the Victor and William Fung Foundation, the Royal Academy of Engineering, EPSRC grants 'Upgrading the small-scale equipment base for early-career researchers in the engineering and physical sciences' and platform grant 'Molecular Quantum Devices' (EP/J015067/1), a Marie Curie Career Integration Grant, and Templeton World Charity Foundation. We thank K. Lui for technical contributions and R. Schouten for discussion.